\documentstyle[prd,aps]{revtex}
\input epsf.sty
\newcommand{\cn}{\mathop{\rm cn}\nolimits}
\newcommand{\tr}{\mathop{\rm tr}\nolimits}
\newcommand{\sn}{\mathop{\rm sn}\nolimits}
\title{ Non-abelian plane waves and stochastic regimes for
(2+1)-dimensional gauge field models with Chern-Simons term}
\author{
D.~Ebert$^{1,}$$^{2}$,
V.~Ch.~Zhukovsky$^{3}$, and
M.~V.~Rogal$^{2,}$$^{3}$}
\address{$^{1}$Research Center for Nuclear Physics (RCNP), Osaka
University, Ibaraki, Osaka 567, Japan}
\address{$^{2}$Institut
f\"ur Physik, Humboldt--Universit\"at zu Berlin, 10115 Berlin, Germany}
\address{$^{3}$Faculty of Physics, Department of Theoretical Physics, Moscow
State University, 119899, Moscow, Russia.}
\date{July 5 2001}
\draft

\begin{document}

\large

\maketitle

\begin{abstract}
An exact time-dependent solution of field equations for
the 3-d gauge field
model with a Chern-Simons (CS) topological mass is found.
Limiting cases of constant
solution and solution with vanishing topological mass
are considered. After Lorentz boost, the found
solution describes a massive nonlinear non-abelian
plane wave. For the more complicate case of gauge
fields with CS mass interacting with a Higgs field,
the stochastic character of motion is demonstrated.
\end{abstract}

\section{Introduction}

The non-abelian Yang-Mills (YM) gauge field theory is essentially
nonlinear. In particular, such characteristic infrared phenomena
of QCD as confinement and chiral symmetry breaking are explained
by the nonlinear nature of gluon interactions leading to the
formation of specific nonperturbative configurations of gauge
fields responsible for these phenomena. Various exact
topologically nontrivial finite energy solutions of the classical
gauge field equations in (3+1)-dimensional space-time with or
without Higgs fields (e.g. instantons\cite{shifman}, vortices
\cite{olesen}, monopoles \cite{hooft}) as well as non-abelian
plane-wave solutions \cite{colem}, \cite{savvidy},
\cite{savvidy1}, \cite{vshivcev} have been found (for a detailed
discussion of YM equations, see, e.g., \cite{zhuk},\cite{rubak}).
In (2+1)-dimensional space-time, a nonlinear topological term (the
so called Chern-Simons (CS) secondary character) which is
responsible for formation of a topological mass of the non-abelian
gauge field, may be added to the field Lagrangian \cite{deser}. In
this case, constant non-abelian solutions of the gauge field
equations with a CS mass were found \cite{peskov}. Clearly, the
investigation of general properties of classical solutions of
gauge field equations is important and may provide new knowledge
about the vacuum structure of a given theory.
Among these properties the irregular stochastic character of the
gauge field dynamics is of great interest (see \cite{book}). The
stochastic behavior of the YM theory has been studied in the
pioneering papers \cite{savvidy}, \cite{savvidy1}, \cite {chir},
\cite{mat}, \cite{savv}. The consideration of possible
interactions with a Higgs field is also interesting, but makes the
problem somewhat more complicated. Investigations of the
stochastic behavior of classical field systems with Higgs fields
but without Chern-Simons term were e.g. made in \cite{vshivcev}.
Moreover, numerical investigations of Poincare sections of the
YM-Higgs theory with an $SU(2)$ doublet Higgs field were performed
in \cite{mat}. Taking only the vacuum expectation value of the
Higgs field into account, the authors of this paper showed
qualitatively the stabilizing role of the Higgs mechanism for the
YM dynamics. Furthermore, in \cite {dey} the dynamical
contribution of Higgs fields was considered. It was shown that a
field system with spatially uniform but time-dependent dynamical
Higgs fields, coupled to a gauge field ($SO(3)$ Georgi-Glashow
model), also demonstrates chaotic behavior with certain threshold
values of the energy.

It is further worth noticing that the chaotic dynamics of the
three-dimensional topologically massive gauge fields (without
Higgs fields) was also studied in the framework of the
time-dependent and spatially-uniform Chern-Simons model in ref.
\cite{gia}. The numerical studies of the CS field equations were
performed there by the Runge-Kutta method. The boundary conditions
were chosen in such a way that all three gauge potentials, now
being considered as coordinates of fictitious ``particles", were
put on a circle. The general motion might thus be interpreted as
that of three effective particles moving in a plane and
interacting nonlinearly through a quartic potential and under the
influence of an ``external magnetic field" of strength $\mu$
($\mu$ is a topological mass). It was demonstrated that a
background ``magnetic field" tries to order the system, forcing
the ``particles" to move in periodic orbits, similar to the
influence of a Higgs condensate on the otherwise chaotic classical
YM dynamics. Therefore, if the strength $\mu$ of the topological
term is large compared to the energy density (in the quantized
theory, this would mean strong coupling), the motion is regular,
while in the opposite case it is chaotic. This ``order to chaos
transition" was observed in numerical studies of the CS-model as
well as in the coupled YM-Higgs model, although the two models
differ in many respects. Note also that the transition from
regularity to irregularity in quantum systems, which are
classically chaotic, was considered for a $d=(2+1)$ YM-Higgs
theory in \cite {hal}.

In the present paper, we consider the generalization of the plane-wave
solutions, found earlier in the $d=3+1$-case (see\cite{colem},
\cite{savvidy}, \cite{savvidy1}, \cite{vshivcev}), to the case of a
(2+1)-dimensional gauge field theory with topological Chern-Simons term.
Clearly, the topological mass provides an additional dimensional
parameter for the
massive nonlinear field configuration in question. In this way, we find
a new plane-wave solution of the corresponding field equations.
Next, the possibility of passing to a constant background field
\cite{peskov} with energy density lower than
the perturbative vacuum energy is considered.
Moreover, we demonstrate that the dynamics of 3-d YM fields with
a CS topological mass interacting with Higgs fields is described
by solutions that, generally speaking, are not regular, but rather
obtain ergodic properties. This study was made for an ansatz which is
specific
for the gauge fields in (2+1)-dimensional space-time and different from
that considered earlier in the (3+1)-d case without CS term or in the
(2+1)-d case with CS term. Our numerical study of the stochastic
properties of solutions with growing energy demonstrates that a chaotic
behavior of solutions is observed mostly for those values of the energy
which are near to the critical ``saddle" point value of the potential
energy of the effective mechanical system. We emphasize that we have
also considered
the important case of spontaneous symmetry breaking in the
CS-Georgi-Glashow model (where the mass term has the sign required for
spontaneous symmetry breaking, and the real mass squared is
negative $-m^2<0$). Finally, the dynamics of a system of
coupled gauge and Higgs fields is investigated in the general case,
where fairly large values of the energy are admitted and, in contrast
to earlier work, no assumptions are made that the system is near a
stable solution (near the ``minimum" critical point of the effective
potential).

\section{Time-dependent solutions}

Consider a 3-dimensional $SU(2)$ gauge field theory with a
Chern-Simons term. The gauge field
Lagrangian can be written as follows
\begin{eqnarray}
{\cal L}_g = -
\frac{1}{4} F_{\mu \nu}^{a} F^{\mu \nu {a}} +
\frac{\mu}{4}\epsilon^{\mu\nu\alpha} [F_{\mu\nu}^a A_{\alpha}^a -
\frac{g}{3}\epsilon^{abc} A_{\mu}^{a}A_{\nu}^{b}A_{\alpha}^{c}] ,
\label{lagr}
\end{eqnarray}
where $F_{\mu\nu}^a =
\partial_{\mu}A_{\nu}^a - \partial_{\nu}A_{\mu}^a +
g\epsilon^{ abc}A_{\mu}^bA_{\nu}^c $
is the gauge field tensor. The last term in (\ref{lagr}) is the
CS term \cite{deser} that describes the
topological mass $\mu$ of the gauge field.

The Lagrangian (\ref{lagr}) is not gauge invariant, since it changes by a full
derivative under a gauge transformation
\begin{eqnarray}
S= \int d^3 x
{\cal L}_g \longrightarrow \int d^3 x {\cal L}_g + \mu (8\pi^2 / g^2)
W (U) ,
\nonumber
\end{eqnarray}
where
\begin{eqnarray}
W (U) = \frac{1}{24\pi^2} \int d^3x \epsilon^{\alpha\beta\gamma} \tr
[\partial_\alpha UU^{-1}\partial_{\beta} UU^{-1}\partial_{\gamma}
UU^{-1}] ,
\nonumber
\end{eqnarray}
and $U(x)=\exp [i\sigma^a\theta^a(x) ]$.
Here $ W (U)$ is the ``winding number" of the gauge transformation
$U(x)$, $\sigma^a$ are Pauli matrices. In order to make $ \exp(i S )$
gauge-invariant, the Chern-Simons parameter $\mu$ should be quantized
\cite{deser}
\begin{eqnarray} \mu= \frac{g^2 n}{4\pi} , n \in Z.
\label{tmass}
\end{eqnarray}
From
the Lagrange equations
\begin{eqnarray}
\frac{\partial {\cal L}_g}{\partial A_{\nu}^a} -
\frac{\partial}{\partial x^{\mu}}\frac{\partial{\cal L}_g}{\partial
A_{\nu , \mu}^a} = 0,
\nonumber
\end{eqnarray}
one obtains the field equation
\begin{eqnarray}
D_{\nu}F^{\nu\mu a} +
\frac {\mu}{2} \epsilon^{\mu\alpha\beta} F_{\alpha\beta}^a = 0 ,
\end{eqnarray}
where $ D_{\mu}F_{\alpha\beta}^{a}=\partial _{\mu} F_{\alpha\beta}^a + g\epsilon^{abc}
A_{\mu}^b F_{\alpha\beta}^c $ .

Let us seek solutions $ A_{\mu}^a = A_{\mu}^a (t)$ that depend
only on the time $t$, but not on coordinates $\bf x$
and choose the following rather restrictive ansatz
\begin{eqnarray}
& A_{\mu}^a = \delta_{\mu}^a f_{a}(t),\ A^{\mu a} = g^{\mu a}f_{a}(t),\\
& a=1,2,3,\ \mu =1,2,3\ (x_1=t , x_2 = x , x_3 = y), & \nonumber\\
& g_{\mu \nu} = {\mbox {diag}}(+,-,-),&\nonumber\\
\end{eqnarray}
where no summation over $a$ is assumed. This case formally corresponds
to a mechanical system with a finite number of degrees of freedom,
whose motion is described by the system of equations
\begin{eqnarray}
\mu g f_2 f_3 + g^2 f_1(f_2^2 +f_3^2) = 0 , \nonumber \\
\mu g f_1 f_3 -\ddot{f_2} + g^2 f_2(f_1^2 -f_3^2) = 0 , \nonumber \\
\mu g f_1 f_2 -\ddot{f_3} + g^2 f_3(f_1^2 -f_2^2) = 0 ,\\
\mu \dot{f_2} + g(\dot{f_1}f_3 + 2f_1\dot{f_3}) = 0 , \nonumber \\
\mu \dot{f_3} + g( \dot{f_1}f_2 + 2f_1\dot{f_2}) = 0 , \nonumber
\end{eqnarray}
\ where $\dot{f_{a}} = \frac{\partial f_{a}}{\partial t}$ . It is
evident that these equations describe three coupled unharmonical
oscillators. It is interesting to note that for $f_1 =f_2 =f_3 =f$
only a trivial solution exists, i.e., $ f =0$. Thus, in order to find
a non-trivial solution, the functions $f_{a}$ should not be taken equal. Let us seek solutions from the class $f_1\ne f_2=f_3=f$. In this case, one obtains
\begin{eqnarray} \mu g f^2 + 2g^2 f_1 f^2 = 0 , \nonumber
\\ \mu g f_1 f -\ddot{f} + g^2 f(f_1^2 -f^2) = 0 , \\ \mu \dot{f} + g(
\dot{f_1}f + 2f_1\dot{f}) = 0. \nonumber \end{eqnarray} \
First of all, we mention that this
system of equations has constant solutions, found earlier in
\cite{peskov}
\begin{eqnarray}
f_1= -\frac{\mu}{2g},\ f_2=f_3=\pm i\frac{\mu}{2g} .
\label{pesk}
\end{eqnarray}
\ Solutions that depend on time can also be found
\begin{eqnarray}
f_1=-\frac{\mu}{2g} ,\nonumber \\
f_2=f_3=f(t).
\end{eqnarray}

Then the corresponding chromoelectric and
chromomagnetic field components are defined as follows
\begin{eqnarray} & E^a_i
= F^a_{1i} ,& \nonumber \\ & E^1_i =(0,0,0),\ E^2_i
=(0,\dot{f},\frac{\mu}{2}f),\ E^3_i = (0,- \frac{\mu}{2}f,\dot{f}) &
\\ & H^a=\frac{1}{2}\epsilon^{ij}F^a_{ij},\ H^1=g f^2,\ H^{2}=H^{3}=0.&
\nonumber
\end{eqnarray}

The function $f(t)$ satisfies the following nonlinear differential
equation
\begin{eqnarray} \ddot{f}+\frac{\mu^2}{4} f+g^2 f^3 = 0 .
\end{eqnarray}
This equation has a first integral which is the energy integral
of motion $\cal E$ for our mechanical system
\begin{eqnarray}
\dot{f}^2+\frac{g^2}{2}f^4+\frac{\mu^2}{4}f^2=const={\cal E}.
\label{ener}
\end{eqnarray}
At the same time ${\cal E}$ is the energy density of the Yang-Mills field
corresponding to the $ T^0_0 $ component of the energy-momentum tensor
of the
field
$$
T^0_0={\cal E} = \frac 12 ( (H^a)^2+(E_i^a)^2).
$$
Note that the Chern-Simons term gives no contribution to the energy density.

The equation (\ref{ener}) can be easily integrated
\begin{eqnarray}
\int \frac {d f} {\sqrt{{\cal E}-\frac{g^2}{2}f^4-\frac{\mu ^2}{4}f^2}}= t-t_0.
\end{eqnarray}
This integral can be expressed in terms of the elliptic functions,
\begin{eqnarray}
f(t) = \frac{\mu}{\sqrt{2} g} \frac{k}{\sqrt{1-2k^2}} \cn
\left( (\frac{\mu^4}{16} +2{\cal E}g^2)^\frac{1}{4} (t-t_0), k\right),
\label{12}
\end{eqnarray}
where $ \cn(x,k)$ is the elliptic Jacobi {\it cosine} function of
the argument $x$ and the module $k$ \cite{korn}
\begin{eqnarray}
k=\sqrt{\frac{1}{2} \bigg
(1-\frac{\mu^2}{4(\frac{\mu^4}{16} + 2{\cal E}g^2)^{\frac{1}{2}}}\bigg)}.
\end{eqnarray} \ This solution is periodical with the period
\begin{eqnarray} T=\frac{4{\rm K}(k)}{({\frac{\mu^4}{16} + 2{\cal
E}g^2})^\frac{1}{4}},
\end{eqnarray}
\ where ${\rm K}(x)$ is a full
elliptic integral of the first kind \cite{korn}.

The oscillation period is inversely proportional to the combination of the
dimensional parameters of the theory, i.e. the CS mass $\mu$ and
the energy of the solution $\cal E$. This, evidently, could be predicted
without solving the equation: there are two scales in the equations,
one of them is provided by a characteristic amplitude of the field
$gA \sim ({\cal E} g^2)^{1/4}$, whose dimension is
$[(gA)] = m \sim 1/T$, and the other is the Chern-Simons parameter with the dimension of a mass $[\mu]=m$.

Since the solution is a function of time only and does not depend on
spatial coordinates, this corresponds to the rest frame of reference
for the field configuration. Therefore, it is quite natural that $ P^i=T^{0i}=0 $, i.e., the energy flux is equal to zero in this frame.

Note that the constant solution (\ref{pesk}) can be obtained
from the solution (\ref {12}) in the limit of an infinite period $ T
\rightarrow \infty $. This is achieved, if we allow ${\cal E}$ to be
equal to $ -\frac{\mu^4}{32g^2}$
\begin{eqnarray} & {\cal E}
\rightarrow -\frac{\mu^4}{32g^2},\ T \rightarrow \infty,\
\frac{k}{\sqrt{1-2k^2}} \rightarrow \pm \frac {i\mu}{\sqrt{2}} &
\nonumber \\ & f(t)\rightarrow \ \pm \frac{i \mu}{2g}. & \end{eqnarray}
It is evident that this limiting procedure can be performed
only if complex values of $f$ are allowed. In this case, the energy
${\cal E}$ remains real, though it becomes negative, which
indicates a possibility of lowering the gauge field energy.

Using our solution, we can pass to the limit of a vanishing
Chern-Simons term corresponding to the Lagrangian
$$
{\cal L}_g = -\frac{1}{4} F_{\mu \nu}^{a} F^{\mu \nu {a}} .
$$
To this end, we assume $\mu \rightarrow 0, A_1^1\rightarrow 0 $ .
As a result, the following limiting solution is received
\begin{eqnarray} A_1^1=0,
A_2^2=A_3^3=(\frac{2{\cal E}}{g^2})^\frac{1}{4} \cn\Big (\ (2{\cal E}
g^2)^\frac{1}{4}(t-t_0), 1/ \sqrt{2}\Big ). \end{eqnarray} \
It should be mentioned that a similar solution was obtained in
ref.\cite{savvidy,savvidy1} for the case of a 4-dimensional
Yang-Mills theory without Chern-Simons term.

\section{Plane-wave solutions}

The solution (\ref {12}) describes a nonlinear standing wave. By applying a Lorentz boost, one can obtain nonlinear propagating waves.
\ It is easy to see that the argument of the above periodical solution
can be Lorentz transformed in the following way:
$ x'_\mu = a_\mu^\nu(\vec v) x_\nu $, $kx = k'x' $, where:
\begin{eqnarray} k'_0 = M \gamma , k'_i =M v_i \gamma, (\gamma
=(1-v^2)^{-1/2}).\end{eqnarray} Here $M$ has the meaning of a mass,
since $k^2= M^2$.

As was already mentioned above, ${\cal E}$ and $g$ have the following
dimensions
\begin{eqnarray} {\cal E}\sim [m]^3,\ g\sim [m]^\frac{1}{2}.
\nonumber
\end{eqnarray}
To define the effective mass of the solution, we have to take the
factor with the dimension of a mass in front of $ (t-t_0) $ in the
cn--function leading to the expression
\begin{eqnarray} M
\sim ({\frac{\mu^4}{16} +2{\cal E}g^2})^\frac{1}{4}. \nonumber
\end{eqnarray}
As expected, the effective mass includes besides the energy density of
the solution the topological mass of the gauge field.
Further investigations of realistic field configurations of this type
should concentrate on the search for possible non-abelian solutions
that describe localized field configurations with a finite energy at
rest (for a general discussion of this point see, e.g., \cite{faddeev}
\cite{novik} \cite{raja}).

\section{(2+1)-d gauge field theory with a CS term and Higgs fields}

In the previous sections we considered the SU(2) gauge field model
in 2+1 dimensions with a Chern-Simons topological term and
obtained non-abelian plane wave solutions. Let us now generalize our
considerations by including in addition a Higgs field contribution
which leads us to a D=3 Georgi-Glashow model with CS
term (``CS-Georgi-Glashow model").

The Lagrangian can now be written as follows
\begin{eqnarray}
{\cal L} = {\cal L}_g+\frac{1}{2}(D_\mu \Phi^a)(D^\mu
\Phi^a) + \frac{m^2}{2}\Phi^a \Phi^a-\frac{\lambda}{4}(\Phi^a \Phi^a)^2,
\label{lagr1}
\end{eqnarray}
where ${\cal L}_g$ is the gauge field Lagrangian (\ref{lagr}), and
$\Phi^a (a=1, 2, 3)$ is the scalar Higgs field in the adjoint
representation ($D_\mu \Phi^a =
\partial_\mu+g\epsilon^{abc}A_\mu^b \Phi^c $). The corresponding
field equations take now the form
\begin{eqnarray}
D_{\nu}F^{\nu\mu a} + \frac {\mu}{2} \epsilon^{\mu\alpha\beta}
F_{\alpha\beta}^a +g\epsilon^{bac}\Phi^c D^\mu \Phi^b = 0 ,\nonumber\\
\ D_\nu D^\nu \Phi^a - m^2\Phi^a+\lambda(\Phi^b\Phi^b)\Phi^a=0 .
\end{eqnarray}

We are again seeking solutions that depend only on the time
$t$, but not on ${\bf x}$, $ A_{\mu}^a = A_{\mu}^a (t),\Phi^a=\Phi^a(t). $
Let us choose the following restrictive Ansatz
\begin{eqnarray}
A_{\mu}^a = \delta_{\mu}^a f_{a}(t),\ A^{\mu a} = g^{\mu a}
f_{a}(t),\ \Phi^a=(\Phi_1(t),\Phi_2(t),\Phi_3(t)),\nonumber\\
\ a=1,2,3,\ \mu =1,2,3\ (x_1=t , x_2 = x , x_3 = y),
\end{eqnarray}
where no summation over $a$ is assumed. Then we arrive at the following
equations of motion for a mechanical system with a finite number
of degrees of freedom
\begin{eqnarray}
\mu g f_2 f_3+g^2 f_1(f_2^2 +f_3^2) +g^2f_1(\Phi_3^2+\Phi_2^2) = 0 ,\nonumber\\
\mu g f_1 f_3 -\ddot{f_2} + g^2 f_2(f_1^2 -f_3^2)-g^2f_2(\Phi_3^2+\Phi_1^2)=0 ,\nonumber\\
\mu g f_1 f_2 -\ddot{f_3} + g^2 f_3(f_1^2 -f_2^2)-g^2f_3(\Phi_2^2+\Phi_1^2) =0,\nonumber\\
\mu \dot{f_2} + g(\dot{f_1}f_3 + 2f_1\dot{f_3})+g^2\Phi_3\Phi_2f_3 = 0 ,\nonumber\\
\mu \dot{f_3} + g( \dot{f_1}f_2 + 2f_1\dot{f_2})-g^2\Phi_2\Phi_3f_2 = 0 ,\nonumber\\
\ g\Phi_1(\dot{\Phi_3}+gf_1\Phi_2)=0,\nonumber\\
g \Phi_1(\dot{\Phi_3}+g \Phi_2 f_1)=0,\\
g \Phi_1(\dot{\Phi_2}+g \Phi_3 f_1)=0,\nonumber\\
g^2 f_2 \Phi_2 \Phi_1=0,\nonumber\\
g^2 f_3 \Phi_3 \Phi_1=0,\nonumber\\
\ddot{\Phi_1}+g^2\Phi_1 (f_2^2+f_3^2)-m^2 \Phi_1+\lambda \Phi^a\Phi^a \Phi_1=0,\nonumber\\
\ddot{\Phi_2}-g(\dot{f_1}\Phi_3+2f_1 \dot{\Phi_3})+g^2 \Phi_2(f_3^2-f_1^2)-m^2\Phi_2+\lambda
\Phi^a\Phi^a \Phi_2=0,\nonumber\\
\ddot{\Phi_3}+g(\dot{f_1}\Phi_2+2f_1 \dot{\Phi_2})+g^2 \Phi_3(f_2^2-f_1^2)-m^2\Phi_3+\lambda
\Phi^a\Phi^a \Phi_3=0,\nonumber
\end{eqnarray}
where $\dot{f_{a}} = \frac{\partial f_{a}}{\partial t},\dot{\Phi^a}=
\frac{\partial \Phi^a}{\partial t} $. Let us seek solutions from the following class $f_1\ne f_2=f_3=f,\Phi_1=\Phi,
\Phi_2=\Phi_3=0$. In this case, one obtains
\begin{eqnarray}
\mu g f^2 + 2g^2 f_1 f^2 = 0 , \nonumber\\
\mu g f_1 f -\ddot{f} + g^2 f(f_1^2 -f^2)-g^2 f \Phi^2 = 0 ,\label{24} \\
\mu \dot{f} + g(\dot{f_1}f + 2f_1\dot{f}) = 0,\nonumber\\
\ddot{\Phi}+2g^2\Phi f^2-m^2\Phi+\lambda\Phi^3=0.\nonumber
\end{eqnarray}
The first equation in (\ref{24}) is solved for $f_1=-\frac{\mu}{2g}$, while
for $\Phi$ and $f$ we get the equations
\begin{eqnarray}
\ddot{f} +f(\frac{\mu^2}{4}+g^2\Phi^2)+g^2f^3=0,\nonumber\\
\ddot{\Phi}+\Phi(2g^2f^2-m^2)+\lambda\Phi^3=0. \label{25}
\end{eqnarray}

\ The potential energy for this system
\begin{eqnarray}
U(f, \Phi)=\frac{g^2}{2}f^4+\lambda \frac{\Phi^4}{4}+ \frac{\mu^2}{4}f^2-
\frac{m^2}{2}\Phi^2+g^2 f^2 \Phi^2
\label{27}
\end{eqnarray}
is presented in Fig. 1.

\begin{figure}
\begin{center}
\epsfxsize=300pt
\epsfysize=240pt

\vspace{5mm}
\parbox{\epsfxsize}{\epsffile{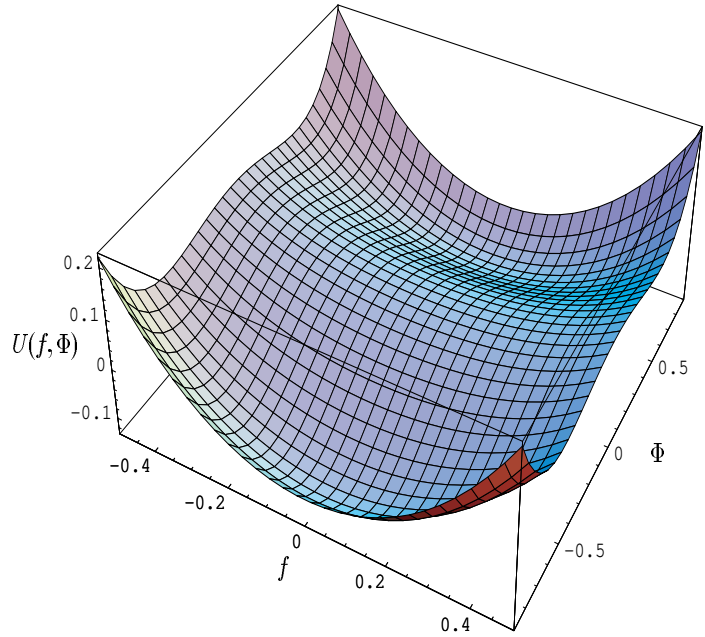}}
\vspace{5mm}
\end{center}
\caption{Potential $U(f,\Phi)$ for $ g=1, \mu=1, m=1, \lambda=2$}
\end{figure}
There are three critical points: two points of the
``minimum" type with the coordinates $ (0, \pm
\sqrt{\frac{m}{\lambda}})$, and one point of the ``saddle" type with
the coordinates $(0,0)$.

The system of equations (\ref{25}) has a first integral, which is the
conserved energy for the mechanical system with coordinates $f$ and $\Phi$
\begin{eqnarray}
\dot{f}^2+\frac{g^2}{2}f^4+\frac{\dot{\Phi}^2}{2}+\lambda \frac{\Phi^4}{4}+
\frac{\mu^2}{4}f^2+g^2 f^2 \Phi^2-\frac{m^2}{2}\Phi^2=const={\cal E}.
\end{eqnarray}
The nonlinear system (\ref{25}) has many solutions, most of them can be obtained
only numerically, and only few can be found analytically. As an example, we present here an analytical periodical solution expressed in terms of
the elliptic Jacobi cosine $ \cn(x,k)$ and sine $\sn(x,k)$ functions of
the argument $x$ and the module $k$
\begin{eqnarray}
f=\frac{1}{g}\sqrt{\left(1-\frac{g^2}{\lambda}\right)\left(\frac{\mu^2}{2}+2m^2\right)}
\cn(Bt,k),\nonumber\\
\Phi=\sqrt{\frac{1}{\lambda}(\frac{\mu^2}{2}+2m^2)} \sn(Bt,k),\label{28} \\
B^2=\frac{1}{4}\mu^2+\left(1-\frac{g^2}{\lambda}\right)
\left(\frac{\mu^2}{2}+2m^2\right),\nonumber\\
k^2=1-\frac{\frac{\mu^2}{2}+m^2}{B^2}.\nonumber
\end{eqnarray}
The energy integral $\cal{E}$ for the solution (\ref{28}) reads
\begin{eqnarray}
{\cal E} =
\left(\frac{1}{g^2}-\frac{1}{\lambda}\right)
\left(\frac{3}{2}m^2\mu^2+\frac{\mu^4}{4}+2m^4\right)+
\frac{\mu^2}{4\lambda}\left(m^2+\frac{1}{4}\mu^2\right).
\end{eqnarray}
It is seen that this solution for $f$ and $\Phi$ is real, if
$\lambda > g^2$.
This evidently describes a regular periodic motion of the system
around the saddle point of the potential, $f=0, \Phi=0$.
The energy $\cal E$ in this case is positive, and well above the
``saddle" point value of the potential $U(f, \Phi)=0$. This classical
solution corresponds to the situation with restored symmetry of the
YM--Higgs Lagrangian in quantum theory (for negative values of the
energy ${\cal E} < 0$
the
effective ``classical" particle should move near
one of the nontrivial minima of the  potential, which corresponds to
broken symmetry of the YM--Higgs Lagrangian).
It is easily seen that with growing $\mu,$ the energy increases
(if $\lambda > g^2$).
Thus, we may suppose that the presence of the CS term is conducive to
restoration of symmetry. For energy values below the saddle point
the solution tends to concentrate near one of the nontrivial minima
of the potential (corresponding to breaking of symmetry). It should
be expected that near the saddle point, trajectories of the system are
highly unstable and the system should demonstrate stochastic behavior.
In order to study this situation we should apply numerical methods of
calculation. This will be done in the following section.

\section{ Effective system with the potential $U(f, \Phi)$}

Here, we will numerically analyze the behavior of the effective
mechanical system
with the potential $U(f, \Phi)$ given in (\ref{27}) leading to
the system of equations (\ref{25}). Let us look for its numerical
solutions for various choices of initial conditions. The functions
$f(t)$ and $\Phi(t)$ for two choices of initial conditions are
plotted in Figs. 2, 3. The curves are depicted for the
following values of parameters $ g=1, \mu=1, m=1, \lambda=2 $.

{
\begin{figure}
\begin{center}
{\epsfxsize=150pt
\epsfysize=150pt
\parbox{\epsfxsize}{\epsffile{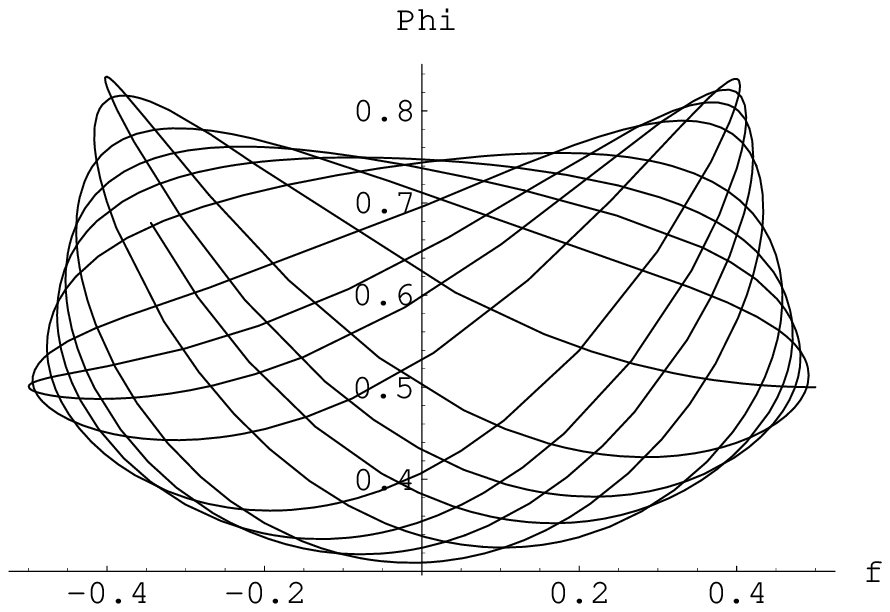}}}
{\epsfxsize=150pt
\epsfysize=150pt
\parbox{\epsfxsize}{\epsffile{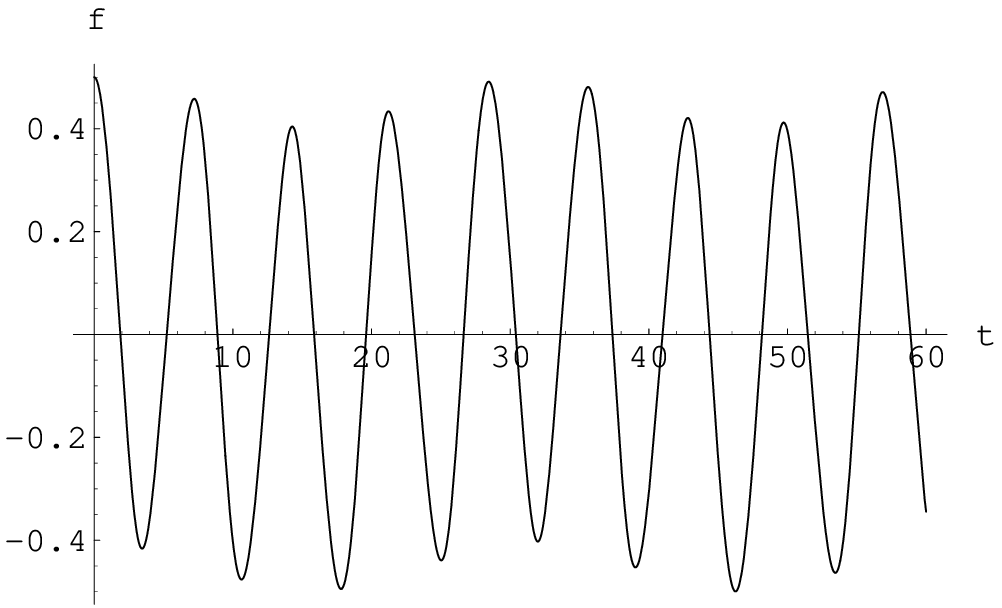}}}
{\epsfxsize=150pt
\epsfysize=150pt
\parbox{\epsfxsize}{\epsffile{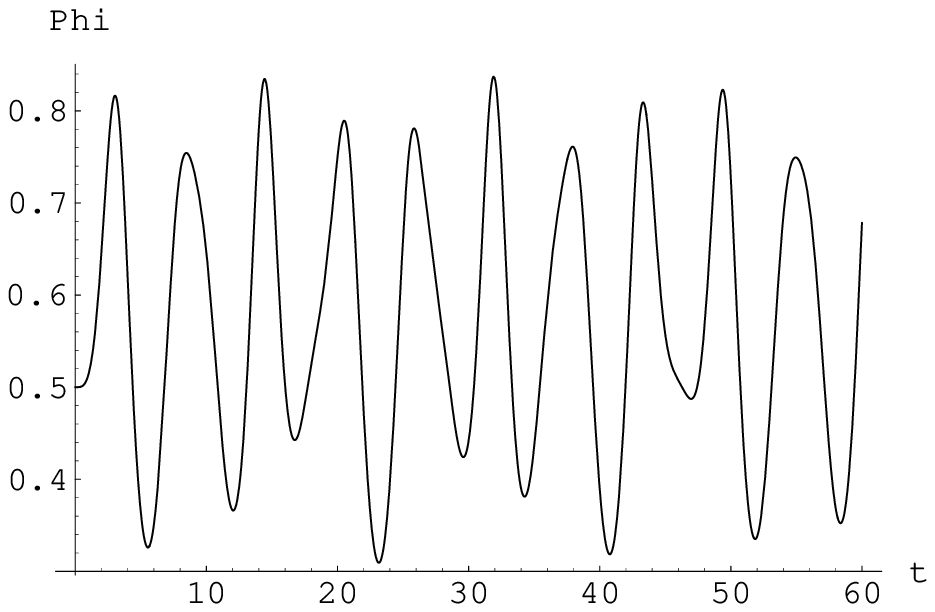}}}
\vspace{5mm}
\end{center}
\caption{System behavior for initial conditions
$f_0=0.5,\dot{f}_0=0,\Phi_0=0.5,\dot{\Phi}_0=0 $ and the set of
parameter values $ g=1, \mu=1, m=1, \lambda=2 $}
\end{figure}
}
From
the figures, one can see that there exist significantly differing
types of motion for the system.
Indeed, the trajectories correspond to three different types of
motion: periodical, quasi-periodical and stochastic ones. The latter situation can be
described by treating the interacting fields as an
effective mechanical system with coordinates $\Phi, f$ and the Lagrangian
\begin{eqnarray} \
L=\dot{f}^2-\frac{\mu^2}{4}f^2-\frac{g^2}{2}f^4+\frac{\dot{\Phi}^2}{2}
+\frac{m^2}{2}\Phi^2-\lambda \frac{\Phi^4}{4}
-g^2 f^2 \Phi^2.
\end{eqnarray}
Let us now consider the term $-g^2 f^2 \Phi^2$ as a perturbation of
the unperturbed system described by the Lagrangian
\begin{eqnarray}
\ L_0=\dot{f}^2- \frac{\mu^2}{4}f^2-\frac{g^2}{2}f^4
+\frac{\dot{\Phi}^2}{2}+\frac{m^2}{2}\Phi^2-\lambda
\frac{\Phi^4}{4}.
\end{eqnarray}
{
\begin{figure}
\begin{center}
{\epsfxsize=150pt
\epsfysize=150pt
\parbox{\epsfxsize}{\epsffile{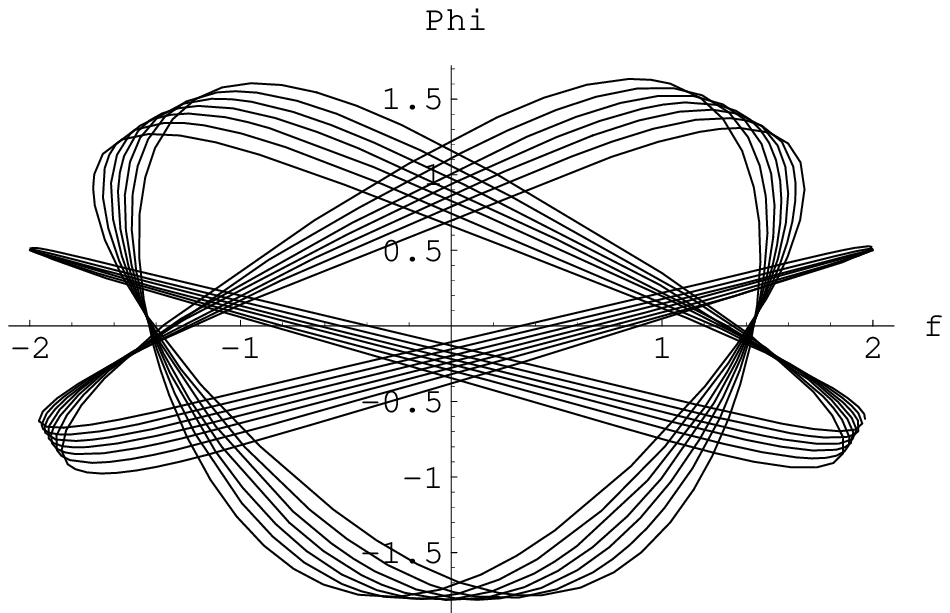}}}
{\epsfxsize=150pt
\epsfysize=150pt
\parbox{\epsfxsize}{\epsffile{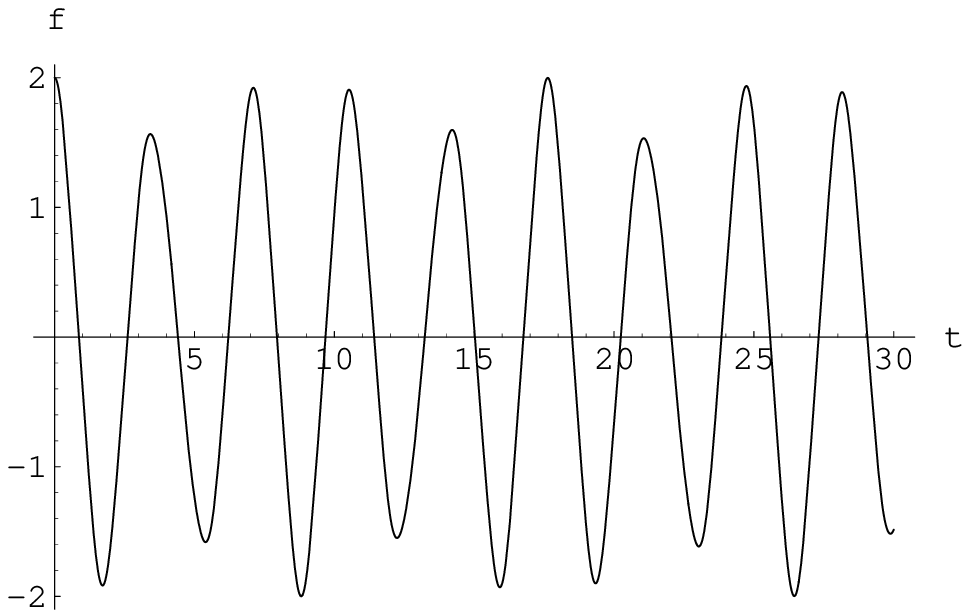}}}
{\epsfxsize=150pt
\epsfysize=150pt
\parbox{\epsfxsize}{\epsffile{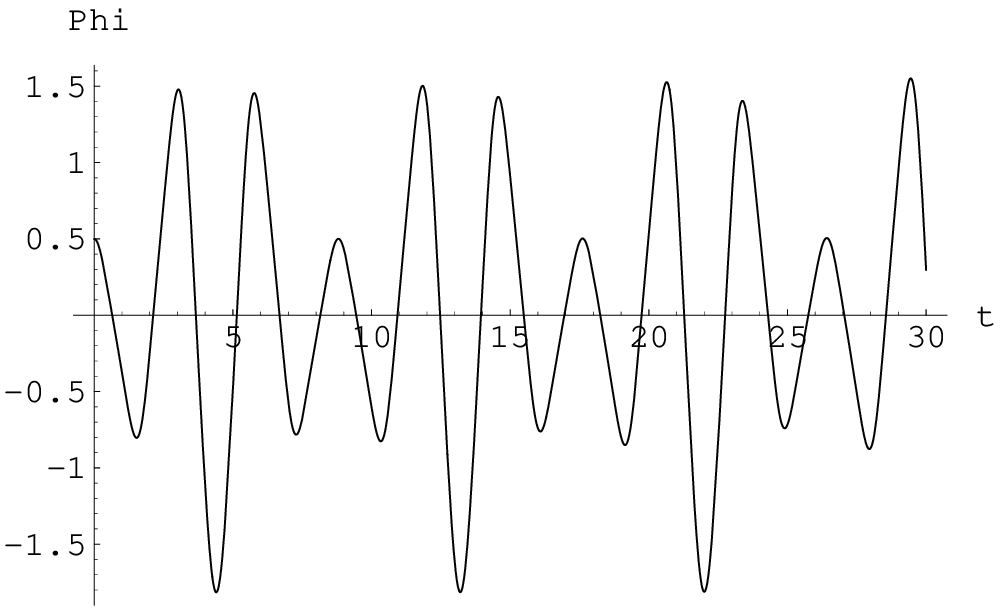}}}
\vspace{5mm}
\end{center}
\caption{System behavior for initial conditions
$f_0=2.0,\dot{f}_0=0,\Phi_0=0.5,\dot{\Phi}_0=0 $ and the set of
parameter values $ g=1, \mu=1, m=1, \lambda=2 $}
\end{figure}
}

For the system with the Lagrangian $L_0$ all trajectories are
either periodical or quasi-periodical, since variables $f$ and
$\Phi$ are separately described by independent equations of the
typical form $\dot{x}^2+a x^2+bx^4=const$ with the solutions $A
\cn(Bx,k)$ and $ A\sn(Bx,k)$ \cite{korn}. In order to describe the
role of the perturbation, the KAM-theorem \cite{kam} can be
applied which states that, if the perturbation is analytical and
small enough, the part of the phase space, spanned by our system,
can be divided into two regions of nonvanishing volumes, one of
them being small as compared to the other and tending to zero in
the limit of a vanishing perturbation. The larger region consists
of invariant embedded toruses, densely covered with trajectories.
In other words, for the majority of possible initial conditions,
trajectories are of the Lissajous type, similar to the oscillator
case. However, there exists a comparatively small region,
corresponding to a small set of initial conditions, where
trajectories are completely chaotic and may go astray from the
neighboring ``restricted" trajectories. This region has a non-zero
volume and it may be called ``instability region".

To examine the trajectories in detail, let us follow \cite{henon}
and seek numerical solutions with the Lagrangian $L$ and
find intersections of trajectories and the plane $f=0$ under the
condition $ f'>0$. Plotting the intersection points in the so
called Poincar\'e surfaces of intersection, we can study the
character of motion for the mechanical system equivalent to our
interacting gauge and Higgs fields. In Fig. 4 we present such
pictures for different values of ${\cal E}$. Each picture contains
a set of trajectories with equal ${\cal E}$. If the set of
intersection points for the given trajectory is restricted to a
point, the motion is periodical; if they lie along the same curve,
the motion is quasi-periodical.
\begin{figure}[h]
\centerline{
{\epsfxsize=95mm
\epsfysize=95mm
\framebox{\epsfbox{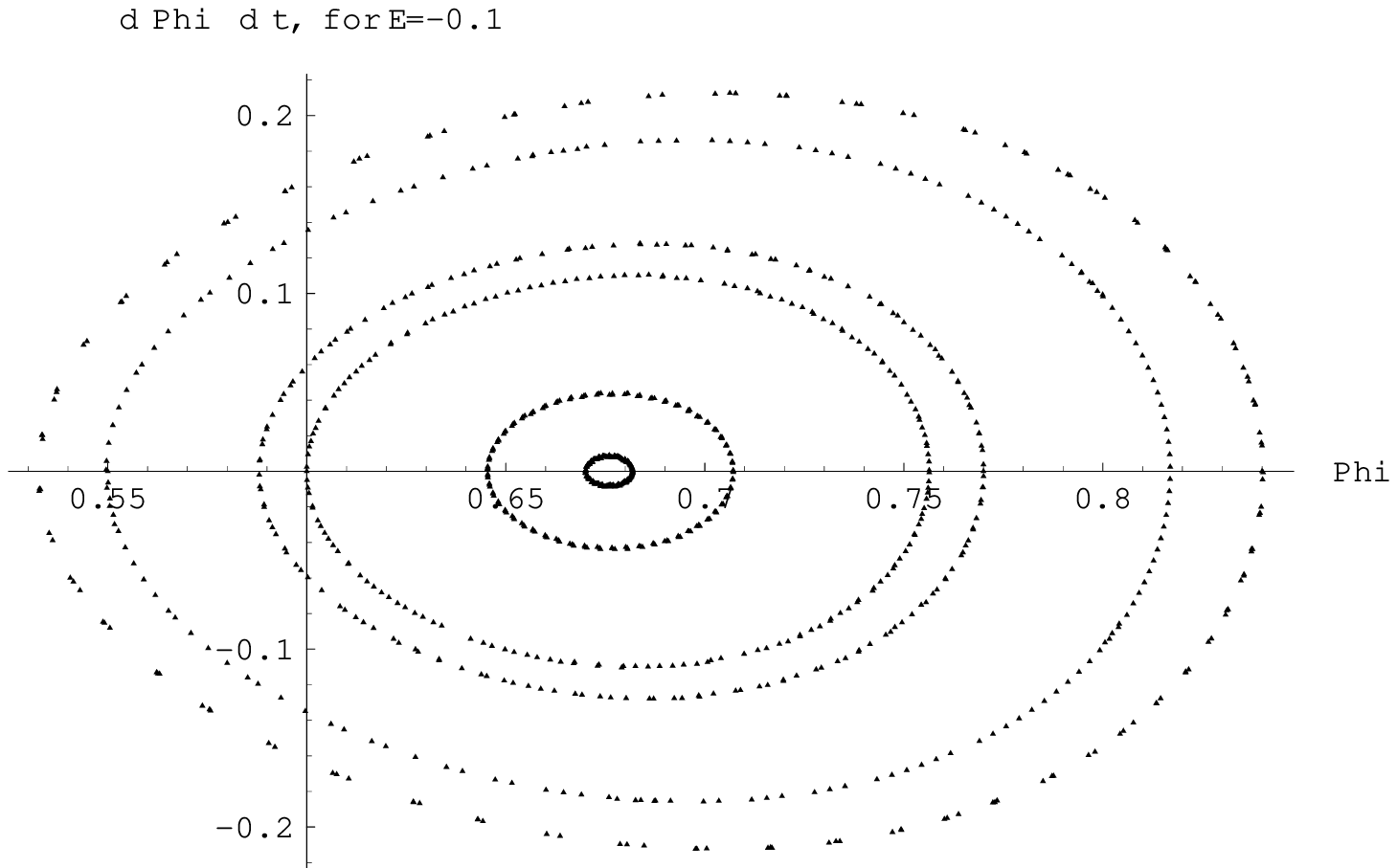}}}
{\epsfxsize=95mm
\epsfysize=95mm
\framebox{\epsfbox{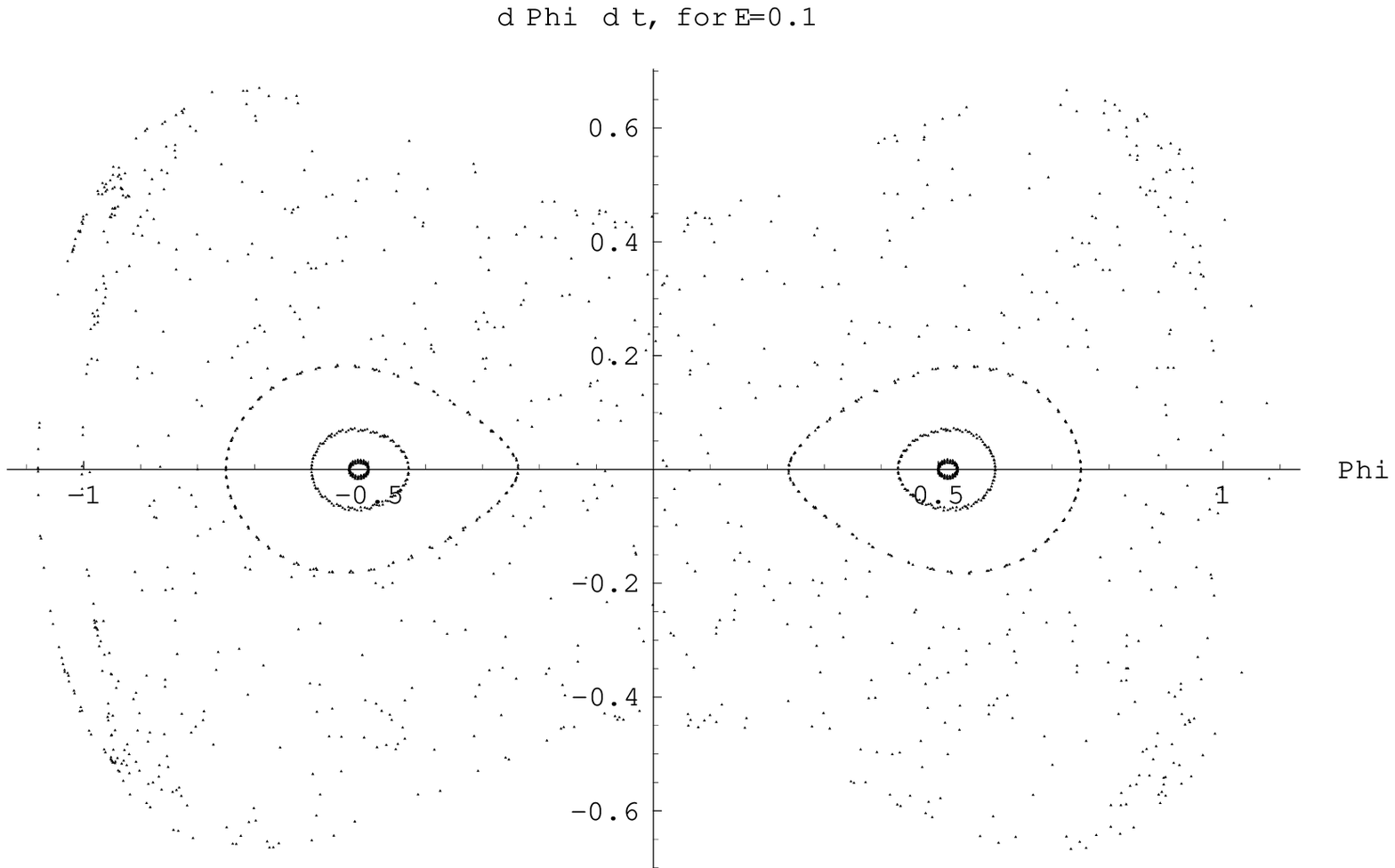}}}
}
\caption{$\Phi$ and $\frac{d \Phi}{d t}$ for $ f=0, \dot{f} > 0, {\cal
E}=-0.1$ and $ 0.1$ ($ g=1, \mu=1, m=1, \lambda=2 $).} \end{figure}

\begin{figure}[l]
\centerline{ {\epsfxsize=95mm \epsfysize=95mm
\framebox{\epsfbox{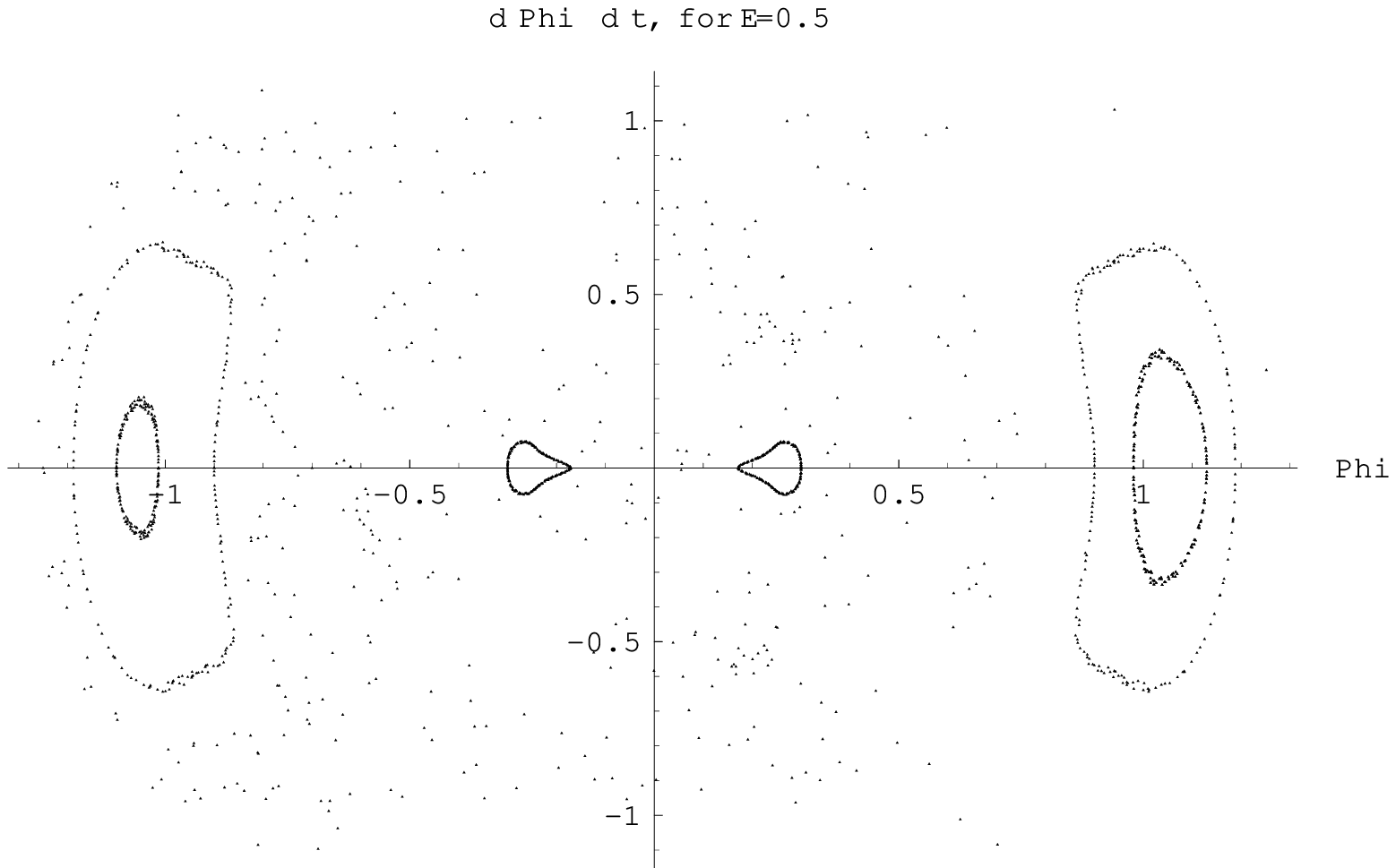}}} {\epsfxsize=95mm \epsfysize=95mm
\framebox{\epsfbox{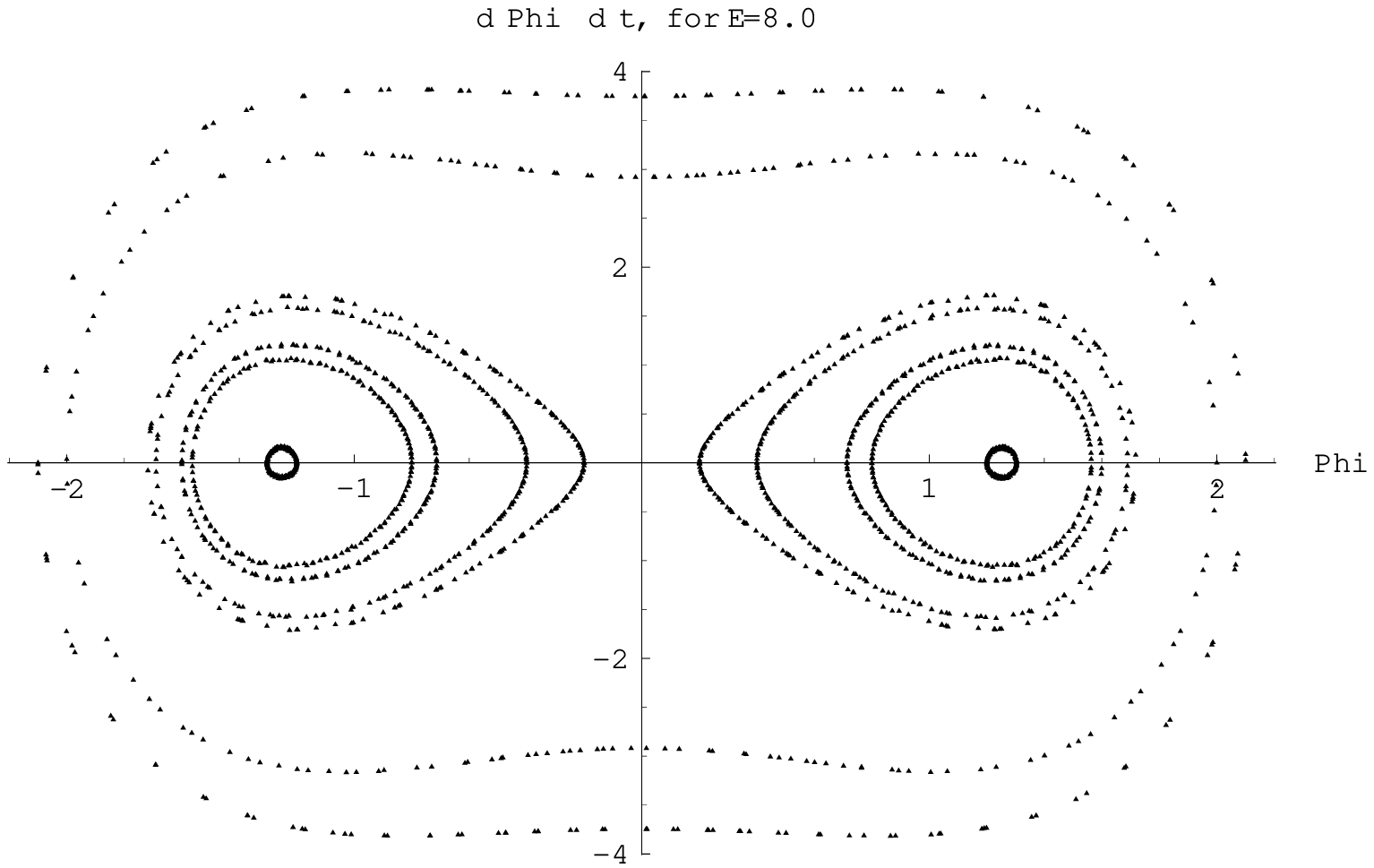}}}
}
\caption{$\Phi$ and $\frac{d \Phi}{d t}$ for $ f=0, \dot{f} > 0, {\cal
E}=0.5$ and $ 8.0$ ($ g=1, \mu=1, m=1, \lambda=2 $).} \end{figure}

For the stochastic motion, the intersection points cover densely a
finite area. The left picture in Fig. 4 corresponds to
comparatively low energies, well below the ``saddle" point. It is
easily seen that all the trajectories form closed curves.
According to KAM-theorem, this means that practically the entire
phase space consists of toruses, which corresponds to
quasiperiodical motion. For higher energies and larger
perturbations (the right picture of Fig. 4), the situation is
completely different: there arise regions of ergodic behavior,
though there remain some small regions of toroidal structure. It
should be clearly understood that dispersed points in Fig. 4 form
one and the same trajectory which chaotically goes through almost
the entire accessible domain of the phase space. For still higher
energies the situation is different. The motion of the system
again becomes more and more stabilized, which is seen in Figs. 4
and 5. The conclusion which follows from these numerical
calculations is that near the saddle point the character of motion
changes, corresponding to the symmetry breaking in the Lagrangian.
The motion becomes more unstable and its stochastic character more
pronounced when the energy is closer to the saddle point value.

We have also made use of the Toda-Brumer criterion \cite{toda} to find
the critical energy $E_c$ such that for energies below $E_c$ no chaotic
trajectories should be found. For our choice of parameters ($ g=1,
\mu=1, m=1, \lambda=2 $)), we found $E_c = -\frac {5} {72}$. This value
does not contradict our conclusion made with regard to numerical
calculations and studies of Poincar\'e surfaces of section (see above).
Unfortunately the Toda-Brumer criterion failed for critical energies
above the saddle point.


\section{Summary and conclusions}

In this paper we have found an exact plane-wave solution of the field
equations for
the 3-d gauge field theory with a non-zero topological mass. The
effective mass of the solution depends on two dimensional parameters
of the model: the amplitude of the wave and the Chern-Simons
topological mass. The plane-wave form of the solution becomes
evident after applying the Lorentz boost. Thus, we obtain a nonlinear
massive plane wave, and this is due not only to the nonlinear
character of the gauge field equations, as was the case with the usual
4-d Yang-Mills equation \cite {savvidy}, but also to the role
of the massive Chern-Simons term. This term, on one hand, adds a
positive contribution to the effective mass of the
solution, and this makes the wave even heavier. On the other
hand, it allows the energy density to become negative. Here, as a
limiting case, we obtained a known static solution with the lowest
energy possible for the class of solutions considered.

We were faced with another interesting situation, when we chose to include
not only the CS term, but also a scalar Higgs field. A choice of the
restrictive ansatz, similar to the one used in the previous Sections,
now led us to an effective mechanical system with two degrees of
freedom. We have demonstrated that
for special initial conditions, the phase space trajectories are closed
and regular. Chaotic behavior of the Yang-Mills theory in the 4-d
space-time was discovered and studied by
Matinyan and Savvidy (see \cite{matin}, and \cite{savv} for reviews
and discussions). In \cite{mat} the $d=2$ dimensional YM-Higgs theory
was studied. There, the Higgs field was chosen to be an $SU(2)$
doublet. In that paper it was qualitatively demonstrated that
spontaneous breakdown of gauge symmetry by the Higgs mechanism has a
stabilizing effect on non-abelian gauge field dynamics. In our case, no
spontaneous breaking of symmetry was initially assumed, rather the
Higgs field was taken as a dynamical variable. Moreover, since we
studied the $d=3$ case, we included the CS term into the consideration.
Our result, based on numerical calculations, is that with growing
energy $E$, the motion of the effective classical mechanical system
first becomes more chaotic (in the vicinity of the critical saddle
point of the potential), then with further growth of energy, again
some regular components may appear.

The general physical conclusion of Matinyan and Savvidy \cite{matin}
is that the Yang-Mills field system is not exactly solvable (otherwise,
the trajectories would have a regular rather than chaotic form). This
conclusion is confirmed in the present paper for the case of a 3-d
gauge field model with a topological CS term and an additional
interaction with a dynamical Higgs field. The presence of the CS
topological mass in the 3-d case is conducive to the restoration of
symmetry and to stabilizing the system.

\section*{Acknowledgements}

One of the authors (D.E.) acknowledges the support provided to him
by the Ministry of Education and Science and Technology of Japan
(Monkasho) for his work at RCNP of Osaka University. He is also
grateful to H. Toki for the kind hospitality at RCNP. Two of the
authors (V.Ch.Zh. and M.Rogal) gratefully acknowledge the
hospitality of Prof. Mueller-Preussker and his colleagues at the
particle theory group of the Humboldt University extended to them
during their stay there. This work is supported in part  by
DFG-Project 436 RUS 113/477/4.


\end{document}